\documentclass[sigconf, preprint, nonacm]{acmart} 

\usepackage{booktabs, multirow, graphicx, enumitem, pifont}
\newcommand{\cmark}{\ding{51}} 
\newcommand{\xmark}{\ding{55}} 
\begin{document}

\title{AFA: Identity-Aware Memory for Preventing Persona Confusion in Multi-User Dialogue}

\author{Mohammad Al-Ratrout, Pavan Ravva, Shayla Sharmin, \\ Aditya Raikwar, 
Ju Young Shin, and Leila Barmaki}
\affiliation{%
  \institution{University of Delaware}
  \country{~}
}
\renewcommand{\shortauthors}{Al-Ratrout et al.}


\begin{abstract}
When multiple people share a single voice assistant, the system conflates their histories: one resident's preferences can leak into another's responses, eroding utility and trust. We call this failure mode \textit{persona confusion}, and we show it is a measurable problem in today's single-user dialogue systems when deployed in shared environments. We present the Adaptive Friend Agent (AFA), a modular framework that combines voice-based speaker identification with per-user memory stores to enable identity-aware, personalized dialogue across multiple users. To support training and evaluation, we construct PAT (Personalized Agent chaT), a synthetic dataset of 58,289 persona-grounded dialogue turns spanning 133 user profiles and 12 real-world scenarios. We evaluate AFA across five LLM back-ends in a standard response-quality benchmark, with a LLaMA-2-70B model fine-tuned on PAT achieving the highest overall performance. To directly measure persona confusion prevention, we introduce an interleaved multi-user evaluation protocol with a novel metric, Persona Attribution Accuracy (PAA), demonstrating that identity-aware routing improves PAA from 35.7\% to 61.3\%. Human evaluation confirms annotators perceive significantly higher personalization in routing-enabled responses. Our results establish that identity-aware user routing is the critical component for preventing persona confusion in multi-user conversational systems. \href{https://anonymous.4open.science/r/AFA-2430/README.md}{Link to our code.}
\end{abstract}



\begin{teaserfigure}
  \centering
\includegraphics[width=.8\textwidth]{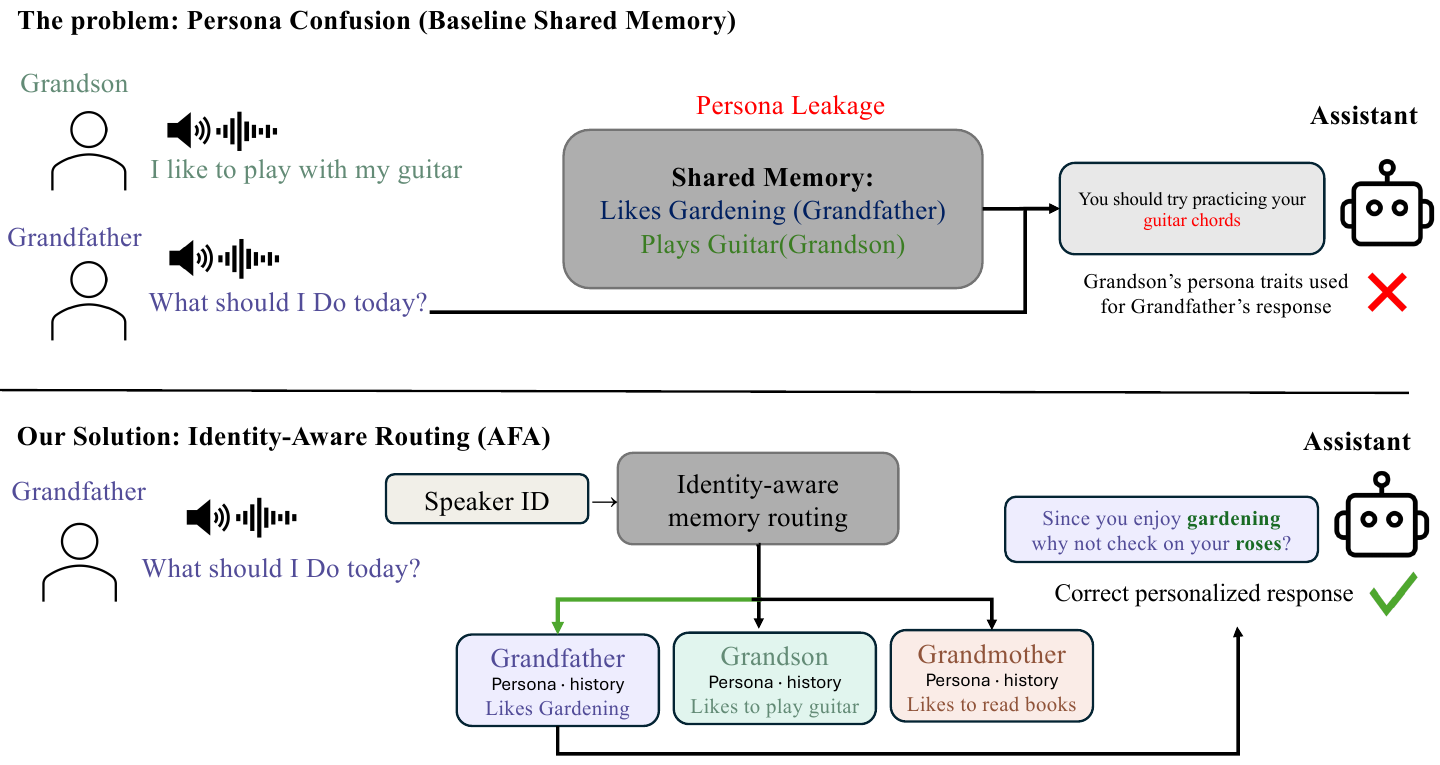}
  \caption{AFA system overview. The Adaptive Friend Agent (AFA) processes spoken audio from multiple residents sharing a single device. Voice embeddings identify who is speaking (Speaker ID), while ASR transcribes what was said. The resolved User ID routes the query to that user's isolated memory store, containing their persona attributes and conversation history, which is injected into the LLM prompt to generate a personalized response. \textbf{Takeaway:} Identity-aware routing is the critical component for preventing persona confusion, improving Persona Attribution Accuracy (PAA) from 35.7\% to 61.3\% over a no-routing baseline.}
  \label{fig:teaser}
\end{teaserfigure}
\maketitle

\section{Introduction}
When two residents share a voice assistant in a communal living environment, the system can confuse them: the grandmother asking about her medication may receive a response shaped by her grandson's earlier gaming questions. We call this failure mode \textit{persona confusion}, and it is not merely a technical error; it is a significant reliability risk. If a shared assistant attributes one resident's social schedule, preferences, or history to another, the system's utility and the user's trust collapse. This risk is amplified in shared residential spaces such as smart homes, communal living facilities, and shared workspaces, where multiple people routinely interact with the same device.

However, existing dialogue agents often assume a single generic user or forget past interactions. Even state-of-the-art chatbots can produce inconsistent answers when users refer back to information from previous sessions~\cite{xu2022beyond}. These inconsistencies underscore the need for explicit long-term memory and user modeling. Prior work on personalized dialogue, such as PersonaChat~\cite{zhang2018personalizing} or multi-session chat~\cite{xu2022beyond}, typically focuses on single-user personalization or fixed persona profiles, creating a gap in handling dynamic multi-user settings.

Our proposed Adaptive Friend Agent (AFA) combines speaker identification with a memory-augmented large language model (LLM), allowing the assistant to retrieve and update each user's personalized memory over time. This integration enables AFA to generate coherent, context-aware, and persona-consistent responses that reflect the evolving preferences of multiple residents across sessions. Our contributions are as follows:
\begin{enumerate}[leftmargin=*]
    \item We develop AFA, a conversational framework that integrates speaker identification, personalized memory retrieval, and LLM-based response generation to support multiple independent users within shared environments.
    \item We construct PAT (Personalized Agent chaT), a large-scale synthetic dataset of 58,289 persona-grounded dialogue turns spanning 133 user profiles and 12 real-world scenarios.
    \item We introduce a novel interleaved multi-user evaluation protocol and the Persona Attribution Accuracy (PAA) metric, which directly measures persona confusion prevention while controlling for topic bias, addressing a gap in existing personalization evaluations that rely on surface-level overlap metrics. Using this protocol alongside human evaluation, we demonstrate that identity-aware routing improves PAA from 35.7\% to 61.3\%, establishing routing as the critical component for multi-user personalization.
\end{enumerate}
\section{Related Work}

\subsection{Multi-User Personalization}
Recent work has highlighted the importance of long-term memory and personalization in dialogue systems. Beyond Goldfish Memory~\cite{xu2022beyond} introduced the Multi-Session Chat (MSC) dataset and showed that standard dialogue models struggle with long contexts spanning multiple sessions. They found that retrieval-augmented and summarization-based models outperform baselines on such data, indicating the importance of explicit memory. While MSC advanced long-term dialogue research, it is limited to repeated sessions between the same pair of speakers and does not address scenarios where multiple distinct users share a single system. This leaves persona confusion, the misattribution of one user's information to another, unexplored. Our work addresses this gap by targeting multi-user personalization with identity-aware routing.
Multi-user AI personalization proposed a multi-agent system workflow to reconcile multiple users' preferences in group settings ~\cite{lee2025map}. We adopt a complementary approach, focusing on a single agent that adapts to each user individually via speaker recognition and per-user memory, rather than coordinating among multiple agents. Moreover, MSC comprises human-human conversational data, while our PAT dataset captures human-AI interactions, which more closely reflects the deployment setting of a conversational assistant.

Concurrent work by Fu et al.~\cite{fu2025pachat} introduces a persona-aware speech assistant for multi-party dialogue (PAChat) that unifies speech and speaker encoding within a fine-tuned LLM architecture. Our work differs in scope and methodology: (1) their system addresses identity-aware response generation within individual conversations, whereas AFA explicitly targets persistent per-user memory across multiple sessions with accumulating conversation history and profile evolution; (2) while their dataset contains multi-party group conversations in single sessions, our evaluation protocol explicitly tests multi-user robustness through interleaved sequences where distinct users alternate turns with the system, directly measuring persona confusion across user switches; and (3) we introduce the Persona Attribution Accuracy (PAA) metric, which directly measures persona confusion prevention by controlling for topic bias, complementing the standard NLG metrics used in prior evaluations.

\subsection{Memory-Augmented Conversational Models}
There is growing interest in equipping LLM-based chatbots with external memory modules to handle long-term context. Notable examples include MemoryBank~\cite{zhong2024memorybank}, which stores and updates key memories of past dialogues, and MemoChat~\cite{lu2023memochat}, which refines LLM instructions to use self-composed ``memos'' for consistent long-range conversation. These methods demonstrate that explicit memory structures can greatly improve response consistency and persona retention over long dialogues. AFA's personalized memory operates on this principle but is specialized for individual users: we maintain a separate memory store for each user and retrieve relevant context when they speak, allowing the assistant to adapt its responses based on that user's history rather than treating all interactions generically.

\subsection{Persona and Profile-Grounded Dialogue}
The role of persona in dialogue has been extensively explored since the introduction of PersonaChat~\cite{zhang2018personalizing}, which demonstrated that conditioning a chatbot on a fixed persona leads to more engaging and coherent interactions. Following this, many datasets such as FoCus~\cite{lee2022focus}, MPChat~\cite{ahn2023mpchat}, and PEC~\cite{zhong2020towards} have investigated persona-based conversations in varying domains.

However, most persona-chat research assumes a static persona profile known in advance and focuses exclusively on single-user settings. Our work similarly leverages initial persona profiles but extends the paradigm to multi-user environments, where the primary challenge shifts from persona quality to persona routing, ensuring the correct profile is applied to the correct user. This approach is conceptually related to retrieval-augmented personalization frameworks such as RAP~\cite{hao2025rap}, which store user-specific information in a structured memory and retrieve it during interaction. While RAP focuses on multimodal assistants that personalize tasks like image captioning and visual question answering, our system extends this paradigm to multi-user, voice-driven dialogue settings with simultaneous personalization across multiple speakers, a capability not addressed in prior single-user or visual-context personalization systems.

\subsection{Speaker Identification}
Speaker identification refers to the process of determining who is speaking from a group of known voices. Voice embeddings are numerical representations of a person's unique vocal characteristics, allowing AI systems to distinguish one speaker from another. Modern approaches use deep neural networks to extract embeddings that capture speaker-discriminative features. ECAPA-TDNN~\cite{desplanques2020ecapa} introduced an architecture combining channel attention, propagation, and aggregation mechanisms, achieving state-of-the-art performance on the VoxCeleb benchmark~\cite{nagrani2020voxceleb}. SpeechBrain~\cite{speechbrainV1} is an open-source speech processing toolkit that provides pre-trained ECAPA-TDNN models for speaker verification and identification. In our framework, speaker identification serves as the entry point for user routing, enabling the system to retrieve the correct user's memory and persona before generating a response. We validate this component's reliability in Section~\ref{sec:speaker_eval}.

\section{PAT Dataset}

We introduce the Personalized Agent chaT (PAT) dataset, which contains query-response pairs tailored to different personalities. This dataset supports training and evaluating personalized dialogue agents in multi-user settings.
Table~\ref{tab:dataset_summary} summarizes the dataset statistics.
\begin{table}[t]
\centering
\footnotesize
\caption{PAT dataset summary statistics.}
\label{tab:dataset_summary}
\begin{tabular}{lr}
\toprule
\textbf{Statistic} & \textbf{Value} \\
\midrule
Total examples & 58,289 \\
Unique personas & 133 \\
Average prompt length (words) & 481.71 \\
Std.\ prompt length (words) & 47.84 \\
Average completion length (words) & 76.41 \\
Std.\ completion length (words) & 14.70 \\
\bottomrule
\end{tabular}
\end{table}

\subsection{Data Generation}
\begin{figure*}[t]
    \centering
    \includegraphics[trim=0cm 1.4cm 0cm 0.7cm, clip, width=\textwidth]{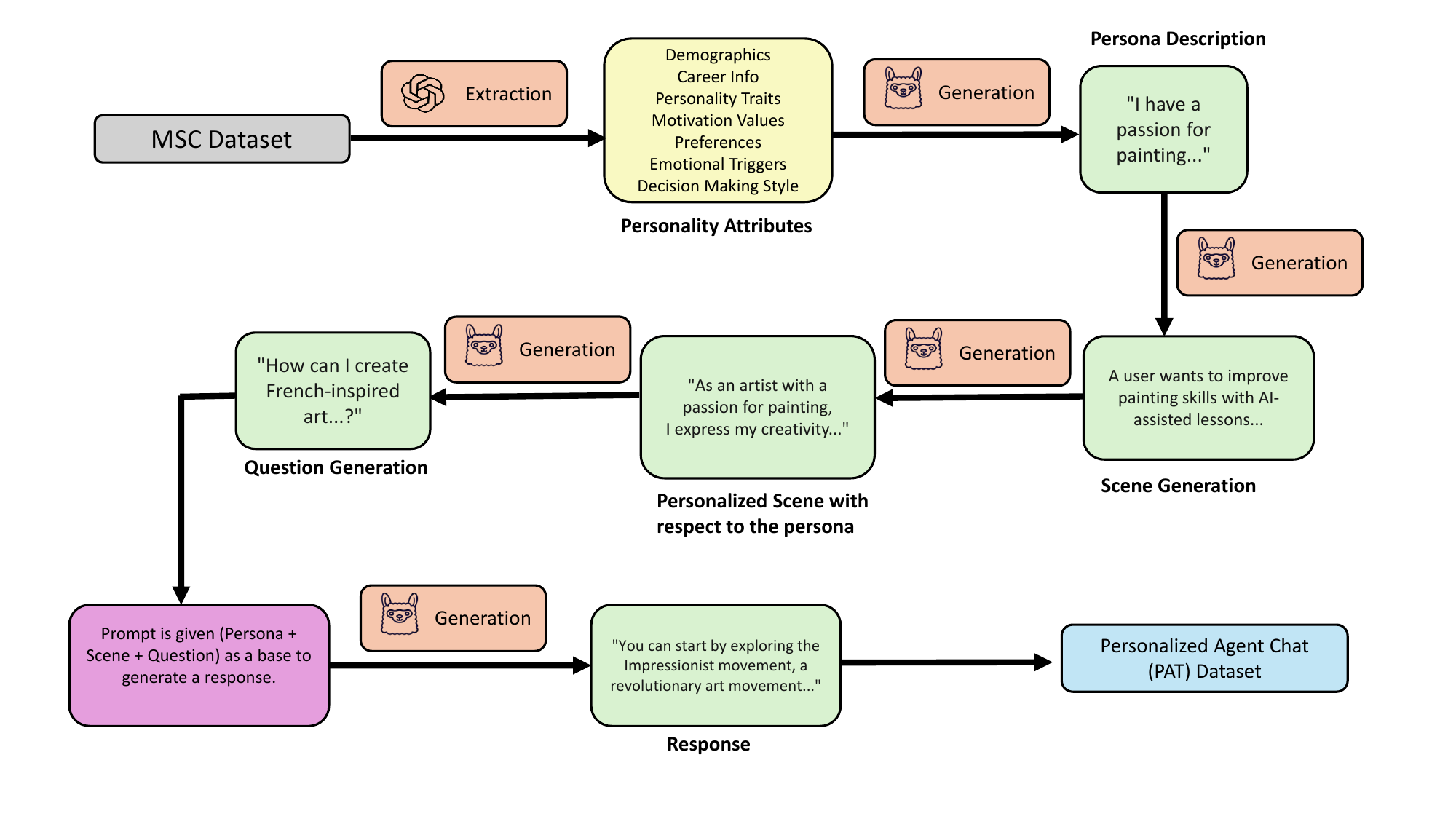}
    \caption{PAT data generation pipeline. Persona attributes and conversational context are extracted from MSC using GPT-4o, then used with LLaMA-405B to generate persona-grounded question-response pairs across 12 real-world scenarios.}
    \label{fig:data_gen}
\end{figure*}

To develop PAT, we used the MSC dataset ~\cite{xu2022beyond} as a source of conversational context. We selected a subset containing coherent multi-turn conversations with distinguishable personal traits, cleaned off-topic segments, and used GPT-4o to extract structured persona profiles. These profiles, combined with scenario contexts, were used to generate persona-grounded human-AI dialogue turns, which better reflect the target use case of AI assistants responding to users.
Figure~\ref{fig:data_gen} illustrates the data generation pipeline.

\subsection{Persona Extraction}
We extracted personality traits from the MSC dataset using GPT-4o, organizing them into six distinct attribute categories:
\begin{enumerate}[leftmargin=*]
    \item \textbf{Demographics}: Nationality, age, gender, and language preference, providing a foundation for understanding the person's background.
    \item \textbf{Career Information}: Educational and career background, enabling assessment of expertise and domain knowledge.
    \item \textbf{Motivations and Values}: Insight into motivations, belief systems, and personal values, enabling the agent to align responses with the user's perspectives.
    \item \textbf{Decision-Making Style}: How an individual arrives at decisions through logical reasoning or emotional intuition.
    \item \textbf{Preferences}: Communication styles, content formats, likes, and dislikes, enhancing user comfort and interactivity.
    \item \textbf{Emotional Triggers}: Emotions and sensitivities that may influence behavior, enabling the agent to avoid responses that cause discomfort.
\end{enumerate}

The extracted personality traits were organized by GPT-4o into cohesive persona descriptions, ensuring all categories are seamlessly integrated into comprehensive, structured representations.

\subsection{Personalized Query Generation}
We generated question-response pairs that mimic real-life user interactions across 12 common scenarios: Project Planning, Language Learning, Job Interview Preparation, Story Development, Hobby Assistance, Personal Development, Emotional Support, Travel Planning, Shopping Assistance, Content Creation, Relationship Advice, and Family Assistance.

Using LLaMA-405B, we generated 40 persona-driven questions per scenario for each persona, structured to be contextually linked so that each successive query builds upon the previous one. Responses were generated by the same model, integrated with a continuously evolving database storing persona descriptions, scene contexts, and prior question-response pairs. Each prompt included the persona description, scenario context, past dialogue, and the current query to ensure continuity. This process produced 58,289 question-response pairs spanning all 133 personas.

\subsection{Dataset Validation}
To evaluate PAT quality, we adopted the LLM-as-a-judge methodology inspired by G-Eval~\cite{liu2023g}, which demonstrated that GPT-4o evaluations strongly correlate with human judgments on NLG tasks. We used GPT-4o~\cite{openai2023gpt4} to rate a stratified random sample of $N = 300$ dialogue samples across four dimensions using a 5-point Likert scale: relevance (4.72), coherence (4.85), fluency (4.99), and helpfulness (4.21). These results, detailed in Table~\ref{tab:llm_eval}, validate the linguistic and structural quality of PAT.

\begin{table}[t]
\centering
\footnotesize
\caption{GPT-4o evaluation of 300 PAT samples (1--5 scale).}
\label{tab:llm_eval}
\begin{tabular}{lcccccc}
\toprule
\textbf{Dimension} & \textbf{Avg} & \textbf{1} & \textbf{2} & \textbf{3} & \textbf{4} & \textbf{5} \\
\midrule
Relevance & 4.72 & 0.3\% & 3.0\% & 2.7\% & 12.7\% & 81.3\% \\
Coherence & 4.85 & 0 & 0.7\% & 2.0\% & 9.0\% & 88.3\% \\
Fluency & 4.99 & 0 & 0 & 0.3\% & 0.3\% & 99.3\% \\
Helpfulness & 4.21 & 0 & 2.7\% & 15.3\% & 40.0\% & 42.0\% \\
\bottomrule
\end{tabular}
\end{table}

\subsection{Illustrative Example}
To illustrate the structure of PAT, Table~\ref{tab:pat-example} presents a representative example. The AI assistant is guided by a rich persona, generating a response consistent with the persona's values, communication style, and contextual needs.
\begin{figure*}[t]
    \centering
    \includegraphics[trim=0cm 13cm 0cm 4.3cm, clip, width=\textwidth]{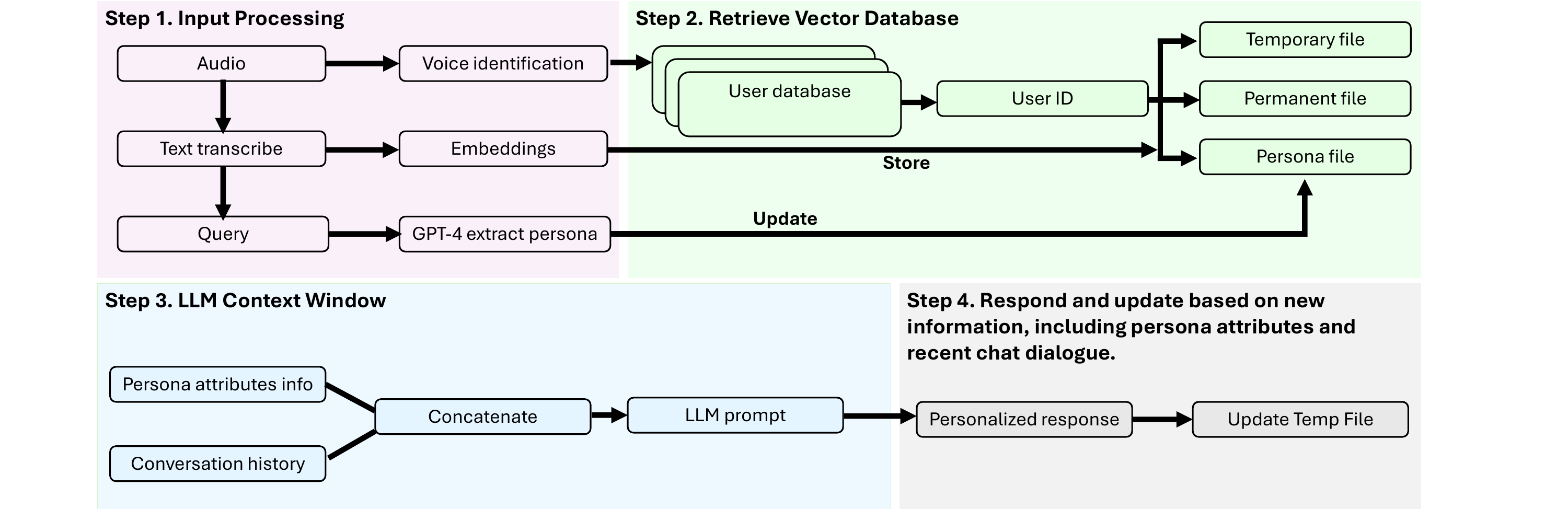}
    \caption{AFA framework overview. The system processes audio input through speaker identification, retrieves user-specific persona and conversation history from a per-user vector database, and generates personalized responses that update the user's profile for adaptive interaction.}
    \label{fig:framework}
\end{figure*}

\begin{table}[t]
\centering
\caption{Example of persona-aligned response generation in PAT.}
\footnotesize
\begin{tabular}{p{0.93\linewidth}}
\toprule
\textbf{Persona Summary:} Works at a non-profit supporting underprivileged youth; values empathy, community, and intellectual growth. Enjoys murder mysteries and ice cream; inspired by Martin Luther King Jr.; prefers collaborative decision-making and tactile experiences like physical books. \\ \midrule
\textbf{User Question:} I'd also appreciate suggestions on where to buy these recommended titles; perhaps there are independent bookstores near me that could use my business. \\ \midrule
\textbf{Generated Response:} You can find these titles at local independent bookstores in your area, which would not only enrich your reading experience but also support the community. To locate these hidden gems, you might check out online directories or ask fellow book lovers for recommendations. By choosing to shop at independent bookstores, you're helping to foster a culture of reading and community engagement that's vital for underprivileged youth. \\
\bottomrule
\end{tabular}
\label{tab:pat-example}
\end{table}

\begin{table}[t]
\centering
\small
\caption{Comparison of persona-grounded dialogue datasets. PAT uniquely combines multi-user testing, human-AI interaction, and multi-session continuity. \textbf{Takeaway:} No prior dataset simultaneously supports all three properties needed for evaluating multi-user personalized assistants over time.}
\label{tab:dataset_comparison}
\begin{tabular}{lccc}
\toprule
\textbf{Dataset} & \textbf{Multi-user} & \textbf{Human-AI} & \textbf{Multi-Session} \\
\midrule
PersonaChat~\cite{zhang2018personalizing} & \xmark & \xmark & \xmark \\
MSC~\cite{xu2022beyond} & \xmark & \xmark & \cmark \\
PAChat~\cite{fu2025pachat} & \cmark & \cmark & \xmark \\
\textbf{PAT (ours)} & \cmark & \cmark & \cmark \\
\bottomrule
\end{tabular}
\end{table}
Compared to PersonaChat~\cite{zhang2018personalizing}, which involves two-human conversations with fixed personas, MSC~\cite{xu2022beyond}, which focuses on multi-session dialogue between the same speaker pair, and the concurrent work of PAChat~\cite{fu2025pachat}, which captures multi-party group conversations within single sessions, PAT focuses on dialogues between an AI assistant and distinct users with diverse persona profiles across multiple scenarios and sessions. Table~\ref{tab:dataset_comparison} summarizes these differences. PAT is the only dataset that simultaneously supports human-AI interaction, multi-user testing, and multi-session continuity, making it uniquely suited for training and evaluating multi-user personalized assistants.
\section{AFA Architecture}

AFA is a modular system that processes voice input, maintains per-user memory, and generates personalized text responses. Figure~\ref{fig:framework} summarizes the pipeline. The key components are described below.

\subsection{Speaker Identification}
The audio data passes through an Audio Identifier Module, which differentiates users based on their unique voice embeddings. We use SpeechBrain's ECAPA-TDNN model~\cite{speechbrainV1,desplanques2020ecapa}, which supports speaker recognition with state-of-the-art performance on VoxCeleb benchmarks~\cite{nagrani2020voxceleb}.

The process begins by converting audio into voice embeddings and transcribing speech to text. These embeddings are stored in a DynamoDB instance, where each embedding is linked with a corresponding user ID. When a user speaks, their voice embeddings are compared against existing embeddings using cosine similarity. If a match is found, the system retrieves the associated user ID; otherwise, a new ID is assigned. This component serves as the entry point for identity-aware routing in the AFA pipeline.

\subsection{Dynamic User Profile Store}
The Dynamic User Profile Store manages historical conversations across multiple sessions for different users. We implement two types of tables for each user: a temporary table storing the last ten conversations for immediate context retrieval, and a permanent table containing summarized session histories.

Once a user completes ten conversations, GPT-4o generates a summary of the session, which is stored permanently while the temporary table is cleared. This mechanism balances short-term context richness with storage efficiency, approximating natural session boundaries similar to a user's daily interaction session. The summarization preserves conversational context and intent while reducing data requirements.

\subsection{Persona Synchronizer}
The Persona Synchronizer updates each user's persona profile after every interaction. After each user turn, GPT-4o is prompted to extract any new personality attributes from the user's query across the six categories described in Section~3.2. The extracted attributes are then compared against the user's existing persona profile: new information, such as previously unmentioned hobbies or preferences, is appended to the profile, while contradictory information, such as a changed job or updated preferences, replaces the corresponding existing entry. The updated persona is stored in the user's profile and used to condition the next response. In the constant persona setting, this update step is skipped, and the persona remains as initially established during onboarding. In the adaptive setting, the persona evolves continuously, enabling the system to reflect changes in the user's preferences over extended deployments.

\subsection{Adaptive Response Generator}
The Adaptive Response Generator integrates multiple components to produce personalized responses. The transcribed query text is processed by both the Dynamic User Profile Store and the Persona Synchronizer. Query text embeddings are compared against stored database embeddings using cosine similarity, and the top-3 most relevant historical data points are extracted. Combined with the user persona and current query, this information is structured into a contextual prompt that ensures the generated response aligns with the user's personality, motivations, and interaction history.

\section{Experiments}

We evaluate the AFA framework through four complementary assessments: speaker identification validation, response quality benchmarking across multiple LLM back-ends, a novel multi-user robustness evaluation, and human evaluation.

\subsection{Speaker Identification Validation}
\label{sec:speaker_eval}

Before evaluating the full framework, we validated the reliability of the speaker identification component that serves as the entry point for user routing. We generated synthetic voice profiles for five speakers (three male, two female) using ElevenLabs text-to-speech, with utterances drawn from the PAT dataset. Each speaker was enrolled with five utterances and tested on ten held-out utterances using SpeechBrain's ECAPA-TDNN model~\cite{speechbrainV1,desplanques2020ecapa}. The system achieved 100\% identification accuracy across all 50 test samples with an average cosine similarity of 0.858, confirming reliable user routing for downstream persona retrieval.


\subsection{Response Quality Evaluation}
\label{sec:response_quality}

We evaluated AFA with multiple LLM backends to assess the framework's generalizability. For open-source models, we used LLaMA-2-70B~\cite{touvron2023llama} fine-tuned on the PAT dataset using LoRA~\cite{hu2021lora} with rank 8, $\alpha$ of 32, and dropout of 0.05, trained over 2 epochs with 8-bit quantization on AWS SageMaker. For closed-source models, we used GPT-4o, GPT-3.5, Claude, and Gemini-2.0 in zero-shot settings.

Each model was tested under three persona configurations: \textit{no persona} (response based solely on dialogue history), \textit{constant persona} (fixed profile established during onboarding), and \textit{adaptive persona} (profile that evolves dynamically based on user interactions).

We measured similarity to ground truth responses using BLEU~\cite{papineni2002bleu} and ROUGE~\cite{lin2004rouge}, diversity using Distinct-1, and persona alignment using Profile-Level and Attribute-Level coverage (P-Cover, A-Cover)~\cite{lin2020xpersona,cheng2024autopal}.

For a persona $p$ containing $n$ attributes $\{a_1, a_2, \dots, a_n\}$ and a generated response $y$, attribute-level coverage is defined as:
\begin{equation}
\text{A-Cover}(y, p) = \max_{a_j \in p}(\text{IDF-O}(y, a_j))
\end{equation}
where IDF-O represents the inverse document frequency weighted word overlap. Profile-level coverage aggregates across responses $S_r$:
\begin{equation}
\text{P-Cover}(S_r, p) = \text{IDF-O}(S_r, p)
\end{equation}

For retrieval, we used OpenAI's text embedding model to encode historical conversations, applying cosine similarity to select the top-3 most relevant entries. We also experimented with top-5 and top-8 retrieval, but observed no significant improvement in response generation quality, so top-3 was selected as the optimal setting.

Table~\ref{tab:persona_results} presents the results. The fine-tuned LLaMA-70B achieved the highest BLEU-1 (0.806) and ROUGE-1 (0.512) under the adaptive persona setting. Similarly, GPT-4o saw a BLEU-1 increase from 0.748 to 0.798 ($\sim$6.7\%) with persona adaptation, demonstrating that persona conditioning enhances response specificity. Persona adaptation quality, assessed through P-Cover and A-Cover, showed the fine-tuned LLaMA-70B achieving the highest values (P-Cover: 0.463, A-Cover: 0.425), indicating strong alignment between generated responses and persona attributes.

The PAT dataset played a critical role in enabling the relatively lightweight LLaMA-70B model to exhibit strong persona adaptation. By providing diverse, well-structured persona-query-response pairs, PAT helped LLaMA-70B approach the performance of larger proprietary models on key personalization dimensions.

This evaluation measures individual response quality by comparing generated responses against ground truth from the PAT dataset. It does not test multi-user routing; each response is evaluated independently for a single user. The multi-user robustness evaluation in Section~5.3 directly tests the system's ability to maintain separate user contexts.

\begin{table*}[t]
\centering
\caption{Response quality comparison across LLM backends and persona settings. Best results per column in \textbf{bold}.}
\label{tab:persona_results}
\footnotesize
\begin{tabular}{llcccccccccc}
\toprule
\textbf{Model} & \textbf{Persona} & \textbf{P-Cov} & \textbf{A-Cov} & \textbf{BL-1} & \textbf{BL-2} & \textbf{BL-3} & \textbf{BL-4} & \textbf{RG-1} & \textbf{RG-2} & \textbf{RG-L} & \textbf{Dist-1} \\
\midrule
\multirow{3}{*}{Anthropic Claude}
& w/o Persona & .394 & .304 & .769 & .687 & .584 & .493 & .390 & .102 & .229 & .810 \\
& Constant    & .408 & .363 & .764 & .681 & .578 & .487 & .374 & .103 & .228 & .817 \\
& Adaptive    & .424 & .410 & .743 & .668 & .570 & .481 & .401 & .105 & .236 & .792 \\
\midrule
\multirow{3}{*}{Google Gemini}
& w/o Persona & .352 & .392 & .528 & .475 & .406 & .348 & .273 & .084 & .175 & .863 \\
& Constant    & .331 & .292 & .574 & .514 & .438 & .376 & .277 & .093 & .184 & .854 \\
& Adaptive    & .370 & .352 & .608 & .545 & .469 & .405 & .299 & .098 & .200 & .851 \\
\midrule
\multirow{3}{*}{OpenAI GPT-3.5}
& w/o Persona & .333 & .305 & .546 & .489 & .416 & .355 & .242 & .074 & .159 & .881 \\
& Constant    & .301 & .379 & .497 & .443 & .377 & .322 & .233 & .068 & .153 & .890 \\
& Adaptive    & .353 & .363 & .590 & .527 & .448 & .388 & .256 & .075 & .168 & .867 \\
\midrule
\multirow{3}{*}{OpenAI GPT-4o}
& w/o Persona & .301 & .294 & .748 & .663 & .562 & .478 & .326 & .097 & .207 & .847 \\
& Constant    & .305 & .284 & .777 & .686 & .581 & .492 & .324 & .099 & .213 & .839 \\
& Adaptive    & .374 & .306 & .798 & .705 & .595 & .502 & .335 & .097 & .217 & .831 \\
\midrule
\multirow{3}{*}{LLaMA-70B (Ours)}
& w/o Persona & .420 & .395 & .782 & .713 & .595 & .495 & .415 & .110 & .229 & \textbf{.894} \\
& Constant    & .435 & .406 & .805 & .734 & .651 & .580 & .497 & .245 & .374 & .777 \\
& Adaptive    & \textbf{.463} & \textbf{.425} & \textbf{.806} & \textbf{.736} & \textbf{.656} & \textbf{.586} & \textbf{.512} & \textbf{.267} & \textbf{.364} & .780 \\
\bottomrule
\end{tabular}
\end{table*}

\subsection{Multi-User Robustness Evaluation}

While the response quality evaluation (Section~\ref{sec:response_quality}) measures how well individual responses align with persona attributes, it does not directly test the system's ability to maintain separate user contexts when multiple users interact with the same device. To address this critical gap, we introduce an interleaved multi-user evaluation protocol.

\subsubsection{Evaluation Protocol.}
We constructed interleaved test sequences by selecting 15 pairs of users with distinct personas from the PAT test set. For each pair, we alternated their questions to simulate a shared environment where two residents speak to the same device in sequence: User~A asks a question, then User~B asks a question, then User~A asks a follow-up to their own earlier question, and so on. Each user contributed 10 turns per sequence, producing 300 total evaluation turns.

For each turn, the system must correctly route to the appropriate user's persona and conversation history. We tested three conditions:
\begin{itemize}[leftmargin=*]
    \item \textit{No persona}: No user identification or persona information; all users share a mixed conversation history. This simulates a conventional single-user assistant.
    \item \textit{Constant persona}: Speaker identification correctly routes to each user's profile, which is established once during onboarding (extracted from MSC conversational data using GPT-4o) and remains fixed throughout all interactions. Conversation history is maintained separately per user.
    \item \textit{Adaptive persona}: Identical to the constant persona setting at the start, the same initial profile extracted from MSC data, but the Persona Synchronizer updates the profile after each interaction, enabling it to evolve over time. Conversation history is maintained separately per user.
\end{itemize}
\subsubsection{Metrics.}
We introduce \textit{Persona Attribution Accuracy} (PAA), which measures whether the system's response shifts more toward the correct user's persona than the wrong user's, controlling for topic bias inherent in the question. For each response $y$, we compute the persona lift toward persona $p$ as:
\begin{equation}
\text{lift}(y, p) = \text{sim}(y, p) - \text{sim}(y_{gt}, p)
\end{equation}
where $\text{sim}$ denotes cosine similarity between sentence embeddings (all-MiniLM-L6-v2~\cite{reimers2019sentence}) and $y_{gt}$ is the ground truth response. Subtracting the ground truth similarity controls for topic bias: a question about cooking will naturally produce responses similar to a culinary persona regardless of personalization. A response is correctly attributed when $\text{lift}(y, p_c) > \text{lift}(y, p_w)$, where $p_c$ and $p_w$ are the correct and wrong user's personas, respectively. PAA is the percentage of correctly attributed responses:
\begin{equation}
\text{PAA} = \frac{1}{N}\sum_{i=1}^{N} \mathbb{1}[\text{lift}(y_i, p_c) > \text{lift}(y_i, p_w)]
\end{equation}
We also report \textit{Persona Margin}, the mean signed difference $\text{lift}(y, p_c) - \text{lift}(y, p_w)$, where positive values indicate correct attribution. Additionally, we compute \textit{Semantic A-Cover} using sentence embeddings to measure semantic alignment between responses and persona attributes, complementing the lexical-overlap P-Cover and A-Cover metrics.

\subsubsection{Results.}
Table~\ref{tab:interleaved} presents the multi-user evaluation results using GPT-4o as the LLM backend. Identity-aware routing produces a substantial improvement in persona attribution: PAA increases from 35.7\% (no persona) to 60.3\% (constant) and 61.3\% (adaptive), representing a 25-percentage-point gain. Persona Margin shifts from $-0.025$ (indicating drift toward the wrong user's persona) to $+0.018$ and $+0.020$ (indicating correct attribution), a sign reversal that demonstrates the system is actively personalizing toward the correct user. Semantic A-Cover shows a corresponding improvement from 0.371 to 0.425--0.431, confirming that responses become more semantically aligned with the intended user's attributes.

Importantly, BLEU-1 and ROUGE-1 remain comparable across all settings (0.300--0.312 and 0.417--0.429, respectively), indicating that routing and persona conditioning improve personalization without degrading general response quality.

The no-persona setting simulates a conventional single-user assistant deployed in a shared environment: all conversation history is pooled without user distinction, mirroring how conventional single-user dialogue systems behave when multiple people interact with the same device. The resulting PAA of 35.7\%, below the 50\% chance level for binary attribution, indicates that the system actively drifts toward the wrong user's persona. For instance, in one evaluation pair, User~A (a rock music enthusiast) asked about clothing style but received a response referencing ``pastel hair colors and floral-inspired makeup'', details from User~B's earlier conversation about Easter hair dyeing. This cross-user contamination confirms that persona confusion is not merely a theoretical concern but a measurable failure mode that identity-aware routing directly prevents.

The modest difference between constant and adaptive persona settings (60.3\% vs.\ 61.3\% PAA) indicates that in short interaction windows, having an accurate initial profile matters more than persona evolution. In practical deployments, initial profiles can be established during an onboarding process, providing reliable personalization from the first interaction. The adaptive mechanism's primary role is to maintain and enrich these profiles over extended use as user preferences evolve.

\begin{table}[t]
\centering
\small
\caption{Multi-user interleaved evaluation (GPT-4o, 300 turns). PAA = Persona Attribution Accuracy. Margin = signed persona lift difference (positive = correct attribution). \textbf{Takeaway:} Identity-aware routing improves PAA by 25 percentage points without degrading BLEU/ROUGE, showing personalization gain is achieved without response quality trade-off.}
\label{tab:interleaved}
\footnotesize
\begin{tabular}{lccccc}
\toprule
\textbf{Setting} & \textbf{PAA}$\uparrow$ & \textbf{Margin}$\uparrow$ & \textbf{S-ACov}$\uparrow$ & \textbf{BL-1} & \textbf{RG-1} \\
\midrule
No persona & 35.7\% & $-$0.025 & 0.371 & 0.305 & 0.424 \\
Constant & 60.3\% & +0.018 & 0.425 & 0.300 & 0.417 \\
Adaptive & \textbf{61.3\%} & \textbf{+0.020} & \textbf{0.431} & 0.312 & 0.429 \\
\bottomrule
\end{tabular}
\end{table}

\subsection{Qualitative Example of Multi-User Routing}

Table~\ref{tab:multi_user_example} illustrates a representative interleaved scenario from our evaluation. When two users alternate turns on the same device, the speaker identification component determines who is speaking and retrieves that user's persona and conversation history before generating a response. Without this routing, the system treats all turns as coming from the same anonymous user and produces generic responses. With routing, each response draws on the correct user's profile, producing personalization that goes beyond what the question alone reveals.

\begin{table}[t]
\centering
\caption{Interleaved multi-user scenario. Two users alternate turns on a shared device. Without user routing, both responses are generic. With routing, User A's response references their music career and British culture interests, while User B's response connects to their coffee shop, despite neither question mentioning these details.}
\footnotesize
\begin{tabular}{p{0.93\linewidth}}
\toprule
\textbf{User A:} \textit{Canadian musician, passionate about British culture.} \\
\textbf{User B:} \textit{Late 20s, owns a coffee shop, drawn to urban art.} \\
\midrule
\multicolumn{1}{c}{\textbf{--- Turn 15: User A (musician) asks ---}} \\
\textit{What marketing channels should I leverage to promote my music and connect with fans?} \\
\midrule
\textbf{Without routing:} Use social media platforms like Instagram and TikTok to showcase your music. Engage with your audience through live streaming on Twitch or Facebook Live. \\
\midrule
\textbf{With routing:} Tap into platforms like Instagram and TikTok for sharing clips. Engaging with \textit{music communities} on Reddit and connecting with \textit{British music scenes} online can also help build your fanbase. \\
\midrule
\multicolumn{1}{c}{\textbf{-- Turn 16: User B (coffee shop owner) asks ---}} \\
\textit{Could you provide guidance on organizing workshops focused on traditional craftsmanship?} \\
\midrule
\textbf{Without routing:} Collaborate with local artisans who can serve as instructors. Select a suitable venue that accommodates hands-on activities and ensure it has the necessary tools. \\
\midrule
\textbf{With routing:} Reach out to local artisans, inviting them to host workshops \textit{in your coffee shop}. This supports the community and enriches \textit{your customers' experiences} by offering hands-on learning. \\
\bottomrule
\end{tabular}

\label{tab:multi_user_example}
\end{table}
\subsection{Human Evaluation}

To validate that automatic metrics correspond to human perception of personalization, we conducted a human evaluation on a subset of the interleaved results. Two annotators independently rated 50 randomly sampled responses, 25 from the no-persona setting and 25 from the adaptive-persona setting, on a personalization scale of 1 (generic, could be for anyone) to 5 (deeply personalized). To mitigate bias, annotators were shown the user's profile, the question, and the response, without knowledge of which setting produced each response, and the presentation order was randomized. Annotators were instructed to assess whether the response connected to the user's profile beyond what the question directly stated.

\begin{table}[t]
\centering
\small
\caption{Human evaluation of personalization (1--5 Likert scale). $N$=2 annotators, 50 items total.}
\label{tab:human_eval}
\begin{tabular}{lcc}
\toprule
\textbf{Setting} & \textbf{Personalization} & \textbf{$n$} \\
\midrule
No persona & 2.32 $\pm$ 0.66 & 25 \\
Adaptive persona & 3.70 $\pm$ 1.06 & 25 \\
\midrule
\multicolumn{3}{l}{\textit{Inter-annotator agreement (within 1 pt): 72\%}} \\
\bottomrule
\end{tabular}
\end{table}

As shown in Table~\ref{tab:human_eval}, annotators rated adaptive-persona responses substantially higher on perceived personalization (3.70 vs.\ 2.32 for no-persona, a difference of +1.38 on the 5-point scale; Welch's $t = 5.52$, $p < 0.001$), confirming the pattern observed in automatic metrics. Inter-annotator agreement within one point was 72\%, indicating consistent judgment across raters. This human evaluation validates that PAA and Semantic A-Cover capture meaningful differences in perceived personalization quality.

\section{Discussion}

Our results demonstrate that \textit{identity-aware routing}, correctly identifying who is speaking and retrieving their specific profile, is the critical component for multi-user personalization. The 25-percentage-point improvement in PAA from no-persona (35.7\%) to constant persona (60.3\%) establishes that user routing alone accounts for nearly all of the personalization gain. The additional improvement from constant to adaptive persona (60.3\% to 61.3\%) is marginal, indicating that persona evolution contributes minimally in short interaction windows. This decomposition reveals that the primary bottleneck in multi-user personalization is not the sophistication of the persona model, but whether the system knows who it is talking to.

A natural question is whether the adaptive mechanism could construct a useful persona entirely from conversation, without an initial onboarding profile. We evaluated this cold-start scenario and found that after 20 turns of interaction, persona attribution accuracy reached only 41\%, substantially below the 60\% achieved with an onboarding profile from the first interaction. This confirms that in the target deployment setting of communal living facilities, where resident information is available through intake processes, providing an initial profile is necessary for reliable personalization.

The speaker identification component achieved perfect accuracy on synthetic voice profiles, validating the front-end of the pipeline. The framework's modular design allows this component to be upgraded independently as more robust models become available, or supplemented with additional identification signals such as facial recognition, proximity-based identification, or user login. The PAT dataset further demonstrated its value by enabling the fine-tuned LLaMA-70B to approach or exceed the performance of larger proprietary models on personalization metrics, suggesting that targeted, persona-grounded training data can compensate for model scale in domain-specific applications.

\paragraph{Limitations.}
The current evaluation uses synthetic voices and the LLM-generated PAT dataset, which may introduce distributional differences from natural conversation. Real-world deployment with natural speech and acoustic variability remains an important area for future validation.

\section{Conclusion and Future Work}

We presented AFA, a modular architecture for personalized, identity-aware conversational support in shared multi-user environments such as smart homes, communal living facilities, and shared workspaces. Through comprehensive evaluation spanning speaker identification validation, response quality benchmarking across five LLM backends, a novel multi-user interleaved evaluation protocol, and human assessment, we demonstrated that identity-aware user routing is the critical factor for preventing persona confusion, improving persona attribution accuracy from 35.7\% to 61.3\%.
Future work includes real-world deployment to evaluate performance with natural speech and acoustic variability, investigating multi-user conflict resolution for reconciling contradictory information across users, evaluating graceful degradation under speaker identification errors, and exploring the framework's applicability to other multi-user settings such as educational environments and collaborative workspaces.

\textbf{Safe and Responsible Innovation Statement:}
Deploying conversational agents in shared multi-user environments raises privacy concerns. AFA addresses these through data isolation: per-user vector databases ensure that one user's conversational history and personal information are not accessible when interacting with another user. Speaker recognition serves as a lightweight biometric layer, ensuring that personalized responses are triggered only by the correct individual. This study uses synthetic data (PAT) to avoid privacy risks during development; future real-world deployments will require informed consent from all participants.
While the Persona Synchronizer automatically replaces contradictory information in user profiles, longer-term risks remain. Stable preferences that become outdated without being explicitly contradicted, such as a hobby the user no longer enjoys but never mentions again, may persist indefinitely. Future iterations will incorporate mechanisms for periodic profile review and user-initiated corrections to address this gap.

\bibliographystyle{ACM-Reference-Format}
\bibliography{references}

\end{document}